\documentclass[10pt]{iopart}
\usepackage{cite}

\usepackage{graphicx}
\usepackage{bm}
\usepackage[dvipsnames,table]{xcolor}
\usepackage{epsfig}

\usepackage{amssymb}
\usepackage{footnote}

\usepackage[utf8]{inputenc} 
\newcolumntype{s}{>{\columncolor[HTML]{F2F2F2}} p{3cm}}

\arrayrulecolor[HTML]{000000} 

\newcommand{\beq}{\begin{equation}}  
\newcommand{\eeq}{\end{equation}}  
\newcommand{\beqa}{\begin{eqnarray}}  
\newcommand{\eeqa}{\end{eqnarray}}

\usepackage{xcolor}

\newcommand{\ESRSTM}{Baumann_Paul_science_2015,Natterer_Yang_nature_2017,Yang_Bae_prl_2017,
Choi_Paul_natnano_2017}

\begin{document}

\title{Enhanced  lifetimes of spin chains coupled to  chiral edge states}

\author{F. Delgado$^1$ and J. Fern\'andez-Rossier$^{2,3}$}
\address{$^1$ Instituto de estudios avanzados IUDEA, Departamento de F\'{i}sica, Universidad de La Laguna, C/Astrof\'{i}sico Francisco S\'anchez, s/n. 38203, La Laguna}

\address{$^2$ QuantaLab, International Iberian Nanotechnology Laboratory (INL),
Av. Mestre Jos\'e Veiga, 4715-310 Braga, Portugal}
\address{$^3$ Departamento de F\'{i}sica Aplicada, Universidad de Alicante,  Spain}
\ead{fernando.delgado@ull.edu.es}

\date{\today}

\begin{abstract}

We consider spin relaxation of finite-size spin chains exchanged coupled with a one dimensional (1D)  electron gas at the edge of a Quantum Spin Hall  (QSH) insulator. 
Spin lifetimes can be enhanced due to two independent mechanisms. First, the suppression of spin-flip forward scattering inherent in the spin momentum locking of the QSH edges. Second, the reduction of spin-flip backward scattering  due to destructive interference  of the quasiparticle exchange,  modulated by  $k_F d$, where $d$ is the inter-spin distance and $k_F$ is the Fermi wavenumber of the electron gas.  
  We show that the spin lifetime of  the $S=1/2$ ground state of odd-numbered chains of antiferromagnetically (AFM) coupled $S=1/2$ spins can be increased more than 4 orders of magnitude by properly tuning the product $k_Fd$ and the spin size $N$, in strong contrast with the 1D case. 
Possible physical realizations together with some potential issues are also discussed.
\end{abstract}

\noindent{\it Keywords:\/ spin Hall, decoherence, adatoms, Kondo}


\maketitle

\section{Introduction}
%
%
The study of both individual spins~\cite{ Heinrich_Gupta_science_2004,Hirjibehedin_Lutz_Science_2006,Khajetoorians_Chilian_nature_2010} and their  engineered  structures~\cite{Hirjibehedin_Lutz_Science_2006,Loth_Bergmann_natphys_2010,Khajetoorians_Wiebe_science_2011,
Loth_Baumann_science_2012,Spinelli_Bryant_naturematerials_2014,Yang_Bae_prl_2017} deposited on surfaces    entails the interaction of these surface spins with the underlying electron gas of the substrate.
These systems can be built and probed using an scanning tunnelling microscope (STM). Moreover, recent experiments~\cite{\ESRSTM} show that single spin resonance experiments can be carried out with STM.
   However,  in order to achieve coherent manipulation of the surface spins  it would be necessary to increase the spin lifetime $T_1$  and spin coherence $T_2$  beyond the inverse of the  Rabi coupling, a condition that is not met in these systems.

 The spin lifetime of individual magnetic atoms is controlled by the electronic  properties of the substrate, the surface spin system, and their exchange interaction, encoded in $\rho J$, 
where $J$ is the  Kondo exchange  and $\rho$ is the substrate density of states~\cite{Ternes_njp_2015,Delgado_Rossier_pss_2017,Ternes_pss_2017}.  The enhancement of spin-lifetimes by reduction  of either $\rho$ or $J$ has been  has been explored in several systems,  including heavy metal substrates with strong spin-orbit~\cite{Hermenau_Ternes_arXiv_2017}, superconducting~\cite{Heinrich_Braun_natphys_2013} or semiconducting~\cite{Khajetoorians_Chilian_nature_2010}  surfaces and thin  decoupling layers, 
such as Al$_2$O$_3$ on NiAl~\cite{Heinrich_Gupta_science_2004}, Cu$_2$N on Cu(100)~\cite{Ruggiero_Choi_apl_2007} or MgO/Ag(100)~\cite{Rau_Baumann_scince2014}. Relaxation times for 3d atoms on metals range from  the ps time scale for magnetic atoms directly on  a conducting substrate~\cite{Khajetoorians_Lounis_prl_2011} up to tens of ms~\cite{Paul_Yang_natphys_2017} in the presence of a decoupling layer.   In addition, the role of the symmetry of the adsorption site have been highlighted~\cite{Miyamachi_Schuh_nature_2013,Hubner_Baxevanis_prb_2014,
Donati_Rusponi_science_2016,Marciani_Hubner_prb_2017}.
This has permitted reaching the single atom limit {\em bit}  by depositing Ho atoms on a MgO/Ag(100) surface~\cite{Donati_Rusponi_science_2016,Natterer_Yang_nature_2017}
with relaxation times exceeding hours observed.

Spin arrays offer
 additional routes to control their spin lifetimes, compared to individual spins on surfaces.
 For instance, transitions between the two ground states of magnetically ordered short spin chains,  either ferro or antiferromagnetically coupled, are severely slowed down~\cite{Loth_Baumann_science_2012,Spinelli_Bryant_naturematerials_2014}.   
A drawback of this approach is that it essentially relies on building  larger objects for which decoherence is much faster. Therefore,  larger $T_1$ comes at the expense of a much shorter $T_2$. A second route for engineering the spin lifetimes of spin arrays is to take advantage of the destructive interference in the scattering between substrate quasiparticles and an ordered array of spins~\cite{Delgado_Rossier_prb_2017}. 

More specifically, in a previous work~\cite{Delgado_Rossier_prb_2017} we found that spin relaxation of linear spin arrays interacting with a 1D electron gas has an oscillatory dependence on $k_F d$.  In this case,  the spin relaxation has two contributions attending to the momentum transfer of the quasiparticle, forward and backward scattering, each of which has both spin-flip and spin-conserving processes.  Here we propose
to use the peculiar spin-locking of the quasiparticles at the edges of a Quantum Spin Hall~\cite{Kane_Mele_prl_2005} to enhance the $T_1$ in engineered spin chains.
For the the edge states of the QSH insulator,  spin-conserving quasiparticle  scattering is only possible in the forward channel, and spin-flip scattering requires back-scattering. Therefore, this already reduces half of the scattering channels.  The scope  of this work is to show how
  spin-flip back-scattering can be further reduced, or even fully suppressed, by engineering $k_F d$, either by atomic positioning (changing $d$) or by gating (changing $k_F$).

%
The quantum spin Hall phase was first observed in heterostructures of HgTe/CdTe~\cite{Konig_Wiedmann_science_2007}, following a prediction of Bernevig {\em et al.}~\cite{Bernevig_Hughes_science_2006}.  Strong evidence of QSH phase has been found on heterostructures of InSb/GaAs~\cite{Du_Li_prl_2017}.    However, the QSH edges are buried inside a heterostructure, making it almost impossible to carry out STM manipulation/deposition of magnetic atoms in these systems.   Edge states have been observed  with STM in atomically thin islands of Bi$(111)$~\cite{Drozdov_Alexandradinata_natphys_2014,Schindler_Wang_natphys_2018}, although there is still some controversy about the evidences of spin-locked edge states~\cite{Yang_Miao_prl_2012,Kim_Jin_prb_2014}. 
  Recently, the observation of the quantum spin Hall effect has been reported on a monolayer crystal of WTe$_2$~\cite{Fei_Palomaki_natphys_2017,Tang_Zhang_natphys_2017} showing a robust topological phase up to 100 K~\cite{Wu_Fatemi_science_2018}.

%
The interplay between local moments and itinerant QSH edge quasiparticles has been extensively studied theoretically~\cite{Maciejko_Liu_prl_2009,Lunde_Platero_2012,Zhang_Kane_prl_2013,Tanaka_Furusaki_prl_2011,Costa_Nardelli_arxiv_2018}.  Most of the previous works have focused on the impact of local spins, either magnetic impurities or nuclei, on the  quasiparticle spin relaxation, which has a direct impact on the lack of conductance quantization:  
 in the absence of spin-flip scattering,  the conductance of QSH channels should be $e^2/h$ \cite{Kane_Mele_prl_2005}. 
 Experiments~\cite{Konig_Wiedmann_science_2007,Du_Li_prl_2017,Fei_Palomaki_natphys_2017,Wu_Fatemi_science_2018} very often find smaller conductance values, which motivates the quest of spin-flip back-scattering mechanisms~\cite{Du_Li_prl_2017,Fei_Palomaki_natphys_2017}. There are in addition several proposals for the manipulation of nanomagnets and local magnetic impurities with the spin-transfer torque exerted by the fully spin-polarized current of QSH edges~\cite{Meng_Vishveshwara_prb_2014,Probst_Virtanen_prb_2015,Silvestrov_Recher_prb_2016}. Here we focus on  the relaxation and decoherence of the spin states of magnetic chains, putting special emphasis on the role of scattering interference. 
 
The rest of this manuscript is organized as follows. In Section~\ref{HAMsec} we describe the model Hamiltonian that describe the spin chain, the electron gas hosted at an edge of a QSHI, and their Kondo interaction.  In Section~\ref{RATESsec} we revisit the expressions for the relaxation $T_1^{-1}$ and decoherence $T_2^{-1}$ rates. .
The dissipative dynamics of a $S=1/2$ degree of freedom encoded in the ground state doublet of odd-numbered chain of antiferromagnetically coupled $S=1/2$ spins is discussed in Section~\ref{NeelR}.  
Finally, a thorough discussion of the results and the main conclusions are presented in Section~\ref{conclusions}

\begin{figure}[hbt]
\includegraphics[width=1.\linewidth]{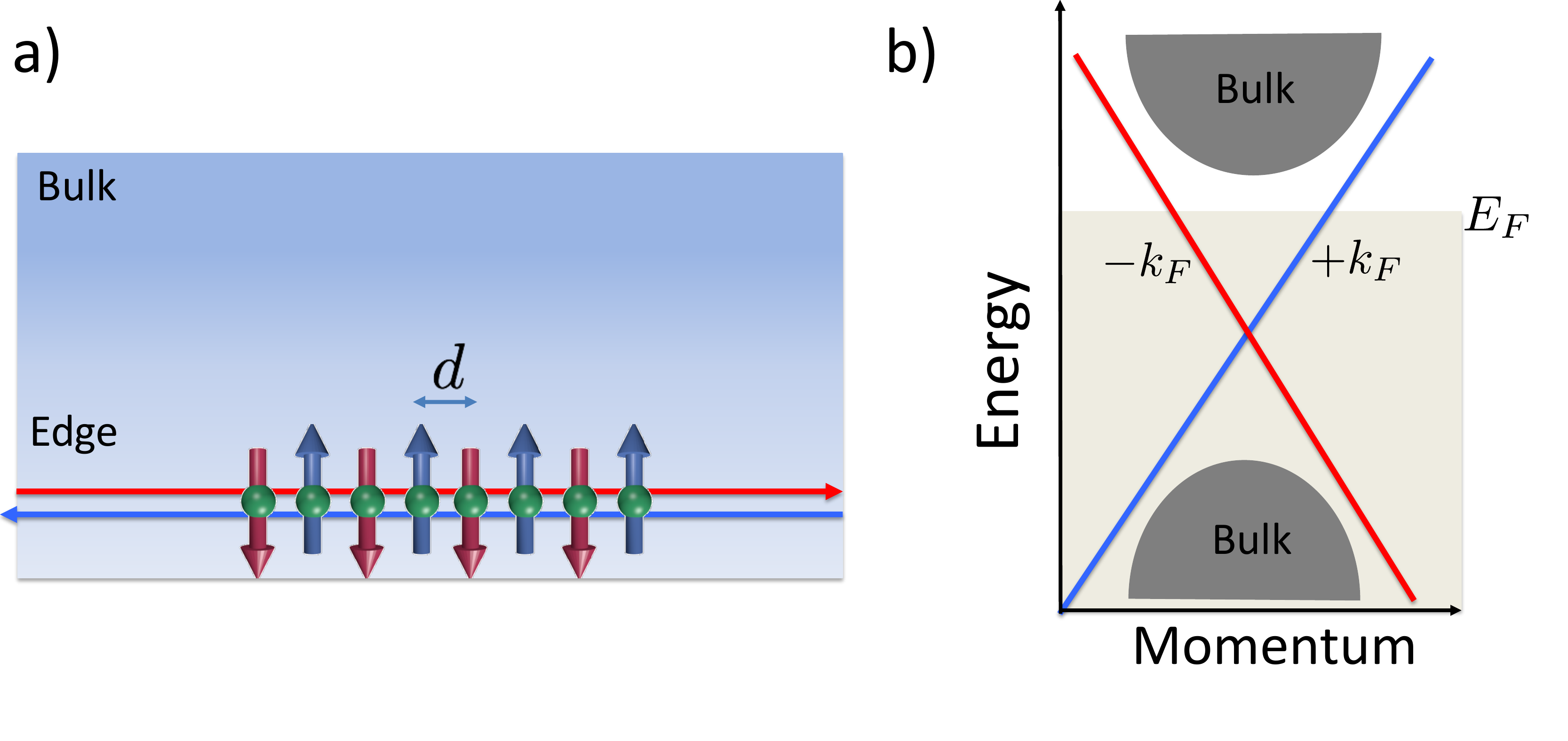}
\caption{ \label{fig0}Scheme of the physical system. (a) A linear antiferromagnetic Heisenberg chain of $S=1/2$ spins is deposited on the edge of an spin Hall bar (SHB). For a given edge, electrons with spin up move in opposite direction to those with spins down.  
(b) Schematic representation of the energy bands of the SHB. The two chiral edge states, for which spin and momentum are locked, present a linear dispersion close to the Fermi level.}
\end{figure}

\section{System Hamiltonian\label{HAMsec}}
Let us consider a linear chain of $N$ equidistant magnetic adatoms, with inter-atom distance $d$, deposited on equivalent atomic adsorption sites described by the spin Hamiltonian ${\cal H}_{\rm chain}=J_H\sum_{l}\vec S(l)\cdot \vec S(l+1)+{\cal H}_{\rm Zeem }$ 
 where $\vec S(l)$ is the spin of the $l$-magnetic adatom. 
 The first term accounts for the Heisenberg exchange coupling and the last one for the Zeeman interaction with an external magnetic field. We will focus our discussion on the simplest $S=1/2$ case,  and thus the magnetocrystalline anisotropy will not be discussed unless otherwise  stated.

%
In the proximity of a conductor, the local spins suffer a Kondo exchange interaction with the conduction electrons~\cite{Misra_book_2007}:
$
{\cal V}_K=  \sum_i^N  {\cal J}_l \vec S(l) \cdot \vec s(\vec r_l),
\label{VKONDO}
$
where $\vec s(\vec r_l)$ is the surface spin density evaluated at the position $\vec r_l$ of the $l$-magnetic atom~\cite{Delgado_Rossier_prb_2017,Delgado_Rossier_pss_2017}.
Here we are interested in the case where the {\em conductor} is substituted by a {\em chiral one-dimensional edge state}.

Spin relaxation of local spins due to Kondo exchange with itinerant quasiparticles is governed by states in the neighbourhood of the Fermi energy. We model these states as left and right moving quasiparticles with linear dispersion $\epsilon_k-\epsilon_F=\pm \hbar v_v k$, where $v_F$ is the Fermi velocity. Taking into account the spin-momentum locking, we define the field operators
\begin{eqnarray}\Psi^{\dagger}_{\uparrow}(x)\approx\frac{e^{ik_F x} }{\sqrt{L}}  \sum_k   e^{ik x} R^{\dagger}_k \nonumber\\
\Psi^{\dagger}_{\downarrow}(x)\approx\frac{e^{-ik_F x} }{\sqrt{L}}  \sum_k   e^{-ik x} L^{\dagger}_k
\end{eqnarray}
where  $L$ is the length of the edge and $L_k^\dag \equiv c_{-(k_F+k),\downarrow}^\dag$ and $R_k^\dag\equiv c_{+(k_F+k),\uparrow}^\dag$ are the left (right) moving fermion operators, 
with $c_{k\sigma}^\dag $ the creation operator 
of a fermion in the edge channel with spin $\sigma$ and momentum $k$. 
\footnote{With this notation, $k>0$ states have $\epsilon_k>\epsilon_F$ both for $L$ and $R$ fermions.}. 
We assume that the quasiparticle spins have a momentum independent quantization axis, 
that we denote as the $z$ axis. This assumption is met  in the Kane-Mele model~\cite{Kane_Mele_prl_2005}, but in real materials it could fail.
We can now write the
 non-interacting Hamiltonian of the QSH edge quasiparticles close to the Fermi energy reads as:
\beqa
{\cal H}_{\rm edge}=\hbar v_F \sum_{k} \left[ (k_F + k) R^{\dagger}_kR_k - (k_F+k) L^{\dagger}_k L_k \right],
\label{Hedge}
\eeqa
where $v_F$ is the Fermi velocity, which is the same for left and right goers on account of time reversal symmetry. Here the sum over $k$ is restrained in a small window around $k=0$ where the linear dispersion relation holds.

A spin chain coupled to one edge state of a SHB can be described
by the following Hamiltonian ${\cal H}={\cal H}_{\rm chain}+{\cal H}_{\rm edge}+{\cal V}_K$, where the resulting Kondo coupling has two different contributions, ${\cal V}_K^\parallel+{\cal V}_K^\perp$, where
\beq
{\cal V}_K^\parallel= \sum_{l}\frac{J_{l}}{2L} \hat S^z(l)\sum_{kk'}
 e^{i(k-k')dl}  \left( R^{\dagger}_k R_{k'} - L^{\dagger}_{-k} L_{-k'} \right)
  \label{VK_par}
\eeq
and
\beq
{\cal V}_K^\perp=
\sum_{l,k,k'}\frac{J_{l}}{2L} \left(\hat S^{+}(l) e^{i(2k_F +k+k')dl} 
L^{\dagger}_k R_{k'} + {\rm h.c.}\right),
 \label{VK_perp}
\eeq
where we have introduced the spin-ladder operators $\hat S^\pm=1/2(\hat S^x\pm i\hat S^y)$.  
The first term ${\cal V}_K^\parallel$ corresponds to the spin-conserving Ising-like term. The second term will be responsible of the spin-flips. Importantly  ${\cal V}_K^\parallel$  contains only forward scattering while ${\cal V}_K^\perp$ accounts for the backward scattering.

\section{Relaxation and decoherence rates\label{RATESsec}}
We describe the dynamics of spin chains coupled to the conduction electrons on the edge states in the SHB within the Bloch-Redfield (BR) approach for open quantum systems weakly coupled to reservoirs~\cite{Breuer_Petruccione_book_2002}. This theory have been applied successfully to describe the inelastic electron spectroscopy~\cite{Loth_Bergmann_natphys_2010,Rossier_prl_2009,Delgado_Palacios_prl_2010,Delgado_Rossier_prb_2010}, renormalization of the magnetic anisotropy~\cite{Oberg_Calvo_natnano_2013,Delgado_Hirjibehedin_sci_2014} 
and the 
dynamics~\cite{Loth_Baumann_science_2012,Spinelli_Bryant_naturematerials_2014,Delgado_Rossier_prb_2010,
Miyamachi_Schuh_nature_2013,
Donati_Rusponi_science_2016,
Delgado_Rossier_pss_2017,Delgado_Rossier_prb_2017}
of magnetic adatoms deposited on a thin decoupling layer.
Here we just applied the general expressions of the transition and decoherence rates presented in Ref.~\cite{Delgado_Rossier_prb_2017} particularized to the coupling terms (\ref{VK_par}-\ref{VK_perp}). For simplicity 
we assume that  the strength of the Kondo interaction $J_l$ is the same for all the atoms in the chain, i.e.,  $J_l=J$.  
After a long but straightforward algebra, one can arrive to the following expressions for the transition rates
$\Gamma_{MM'}$ from state $M$ to $M'$
\beqa
\label{ratesSHI}
\Gamma_{MM'} &=&
\frac{\gamma_S(\omega_{M'M})}{4S^ 2}
\sum_{l,l'} 
\Big[  S^z_{MM'}(l)S^z_{M'M}(l')+
\crcr
&&\hspace{-1.5cm}{\cal F}_{l-l'}( k_F d) \left[S^x_{MM'}(l)S^x_{M'M}(l')+S^y_{MM'}(l)S^y_{M'M}(l')
\right]
\Big]
\crcr
&&
\eeqa
where 
${\cal F}_{l-l'}(\xi) = \cos\left( 2\xi(l-l')\right)$ and
$S_{MM'}^a(l)$ stands for the matrix elements of the $\hat S^a(l)$ spin operator in the $|M\rangle$-eigenbases of $ {\cal H}_{\rm chain}$, and  $\omega_{M'M}=(E_{M'}-E_M)/\hbar$. Here we have introduced the energy and temperature-dependent transition rate of a single spin $S$ exchange coupled to an 
electron gas
\beq
\gamma_S(\omega_{M'M})=\frac{\pi}{2}\left(\rho J\right)^2 S^ 2 {\cal G}(\omega_{M'M})
\eeq
 where ${\cal G}(x)=x/(e^{\beta\hbar x}-1)$, with $\beta=1/k_BT$ and $k_B$ the Boltzmann constant.

There are two kind of scattering processes leading to decoherence. One of them is the so-called {\em non-adiabatic decoherence}, which is associated to population scattering and whose rate is given by~\cite{Breuer_Petruccione_book_2002}:
\beqa
\gamma_{MM'}^{\rm nonad.}=\frac{1}{2}\left(\sum_{N\ne M} \Gamma_{MN}+\sum_{N\ne M'} \Gamma_{M'N}\right).
\label{decohR}
\eeqa
Even if population scattering is strictly forbidden ($\gamma^{\rm nonad.}=0$), there can be a second source of decoherence, the {\em adiabatic decoherence}, associated to the loss of phase coherence due to scattering with the itinerant electrons. 
Following the method outlined in our previous work~\cite{Delgado_Rossier_prb_2017}, we obtain  the adiabatic decoherence rate
\begin{eqnarray}
   \gamma_{MM'}^{\rm ad.}&=&
  \frac{\gamma_S(0)}{8S^2}
  \Bigg[
  \left| \sum_l \left[S^z_{MM}(l)- S^z_{M'M'}(l)\right] \right|^2
  \crcr
  &&\hspace{-1.5cm}+\sum_{b=x,y}
 \left| \sum_{l}\left( e^{i2k_Fdl}S^b_{MM}(l)- e^{-i2k_Fdl}S^{b}_{M'M'}(l)\right)\right|^2 
  \Bigg].
 \crcr
 \label{decoheb}
 &&
\end{eqnarray}
Equations (\ref{ratesSHI}) and (\ref{decoheb}) are the central results of this work.
In deriving these two equations we have assumed a constant spin-independent density of states around the Fermi level, $\rho(\epsilon)=\sum_{k\sigma} \delta\left(\epsilon-\epsilon_{k\sigma}\right)\approx \rho$. In addition, 
we have approximated $k(\epsilon)\approx \pm k_F$~\cite{Delgado_Rossier_prb_2010}.
 Importantly, both $\Gamma_{MM'}$ and $\gamma_{MM'}^{\rm ad.}$ contains a spin conserving contribution (involving only $S^z(l)$) associated to forward scattering, and a spin-flip contribution associated only to backward scattering. 
 
We can always classify quasiparticle scattering, responsible of both $T_1$ and $T_2$,  in terms of the amount of spin exchanged with the spin chain, $\Delta S_z= \pm 1, 0$.  In one dimension, we can fuether split 
  quasiparticle scattering in two groups, according to the variation of quasiparticle momentum, $\Delta k\approx 0, \pm 2k_F$, known as forward and backward scattering respectively.  We have thus 4 types of scattering channels, listed in the table (\ref{Table1}). 
  Importantly, in the case of quasiparticles at the  edge states of the QSH insulators, two of these four channels are forbidden,  
  on account of the spin-momentum locking.  This holds regardless of the properties of the spin chain. The rest of  the discussion shows that,  in addition,  the  spin-flip backward  scattering can also be suppressed in suitably taylored spin chains. 

 \begin{table}[htp]
\caption{Quasiparticle scattering channels, leading to spin relaxation in the spin chains, in 
the edge of  QSH  insulator depending on the variation of the quasiparticle spin
 $\Delta S_z$ and momentum transfer $\Delta k$.} 
\label{Table1}
\begin{center}
\begin{tabular}{ |c|c|c|  }
\hline
 &  $\Delta S_z$ & $\Delta k $ \\
\hline
QSH Allowed  & 0 & 0 \\
\hline
QSH Allowed  & $\pm$ 1 & $2k_F$ \\ \hline
QSH forbidden & 0 & 2 $k_F $ \\ \hline
QSH forbidden  & $\pm$ 1 & 0 \\
\hline
\end{tabular}
\end{center}
\end{table}%

\section{Relaxation and decoherence of the spin-uncompensated N\'eel states \label{NeelR}}
\begin{figure}[t] 
\includegraphics[width=1.1\linewidth]{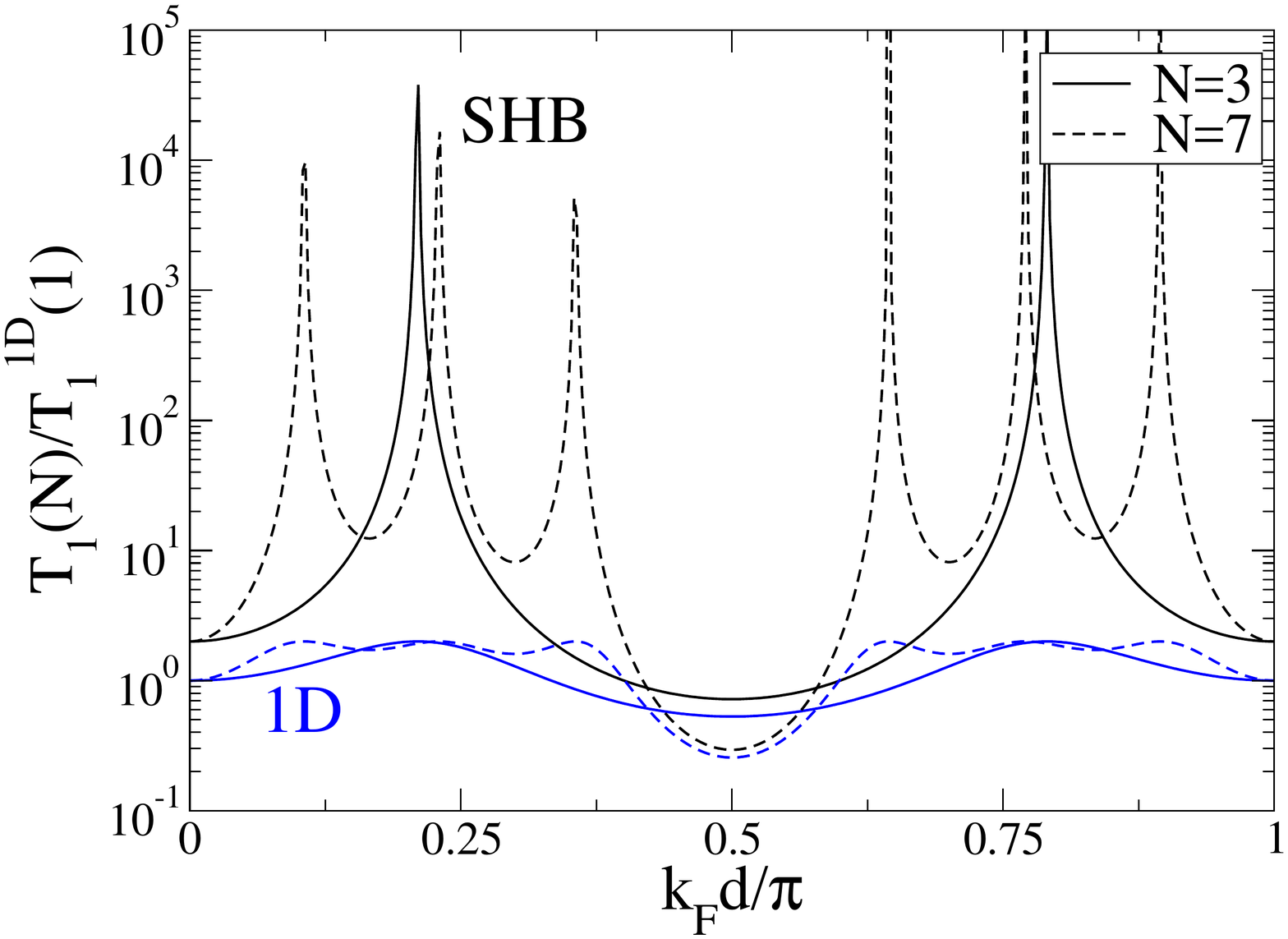}
\caption{ \label{fig3}
Spin relaxation times for odd $N=3$ (solid lines) and $7$ (dashed lines) 
 $S=1/2$ spin chains.  The  relaxation times obtained for the coupling with a SHB (one dimensional conductor) are depicted in black (blue). In all cases, $J_H=1$ meV, $g\mu_B B=0.05$ meV and $k_bT=0.01$ meV. All times are referred to the relaxation time $T_1^{1D}(1)$ of the single spin.}
\end{figure}

Let us consider an antiferromagnetic spin chain with an odd number of $S=1/2$ spins, which yields a
 $S=1/2$ ground state.
The lowest energy states of these chains are staggered N\'eel-like states, where $S_{M,M}^z(l)\sim(-1)^{l+M} {\cal S}(l)$, with ${\cal S }(l)\sim 1/2$.
 Figure \ref{fig3} shows the relaxation time $T_1(N)$ in units of the lifetime of a single spin coupled to a one-dimensional electron gas, $T_1^{\rm 1D}=1/\gamma_{1/2}(\Delta/\hbar)$, where $\Delta=g\mu_B B$.
For completeness, we compare two chains with $N=3,7$ spins coupled to either a spin-locked electron gas, labelled with SHB, or to a normal one dimensional conductor. In all cases, $T_1(N)$  are periodic functions of $k_Fd$, with period $\pi$. We find that, for all values of $k_F d$,  $T_1(N)$ is larger for the spin-locked case than the normal 1D electron gas.  This can be understood right away because, for the $S=1/2$ doublet, $T_1(N)$ entails a flip-flop spin interaction between the quasiparticle bath and the chain state.  In the spin-locked edge states case,  this is only allowed in the back-scattering channel, see equation (\ref{ratesSHI}).

 The most striking feature in figure \ref{fig3} is the very large modulation of $T_1$ as a function of $k_F d$,  which is dramatically larger for the spin-locked reservoir. By choosing the right values of $k_Fd$, one can increase the relaxation time $T_1$ by more than 4 orders of magnitudes. This control could be achieved either by modifying the inter-spin distance, 
 or by a gate voltage that could vary the Fermi level of the electrode.  Notice that, from a practical point of view, changing either $d$ or $k_F$ will probably change the intrachain exchange $J$, an effect that will not be discussed here.
  Notice that spin chains of length $N$ presents $N-1$ maxima of $T_1(N)$ in each period of $k_Fd$, with the maxima found when $(k_Fd)^{-1}\sim 2(N-1)/\pi$. These maxima in $T_1$  can be attributed to the suppression of spin-flip back-scattering due to destructive interference in the multiple back-scattering of the quasiparticle with the spin chain.

\begin{figure}[t]
\includegraphics[width=1.1\linewidth]{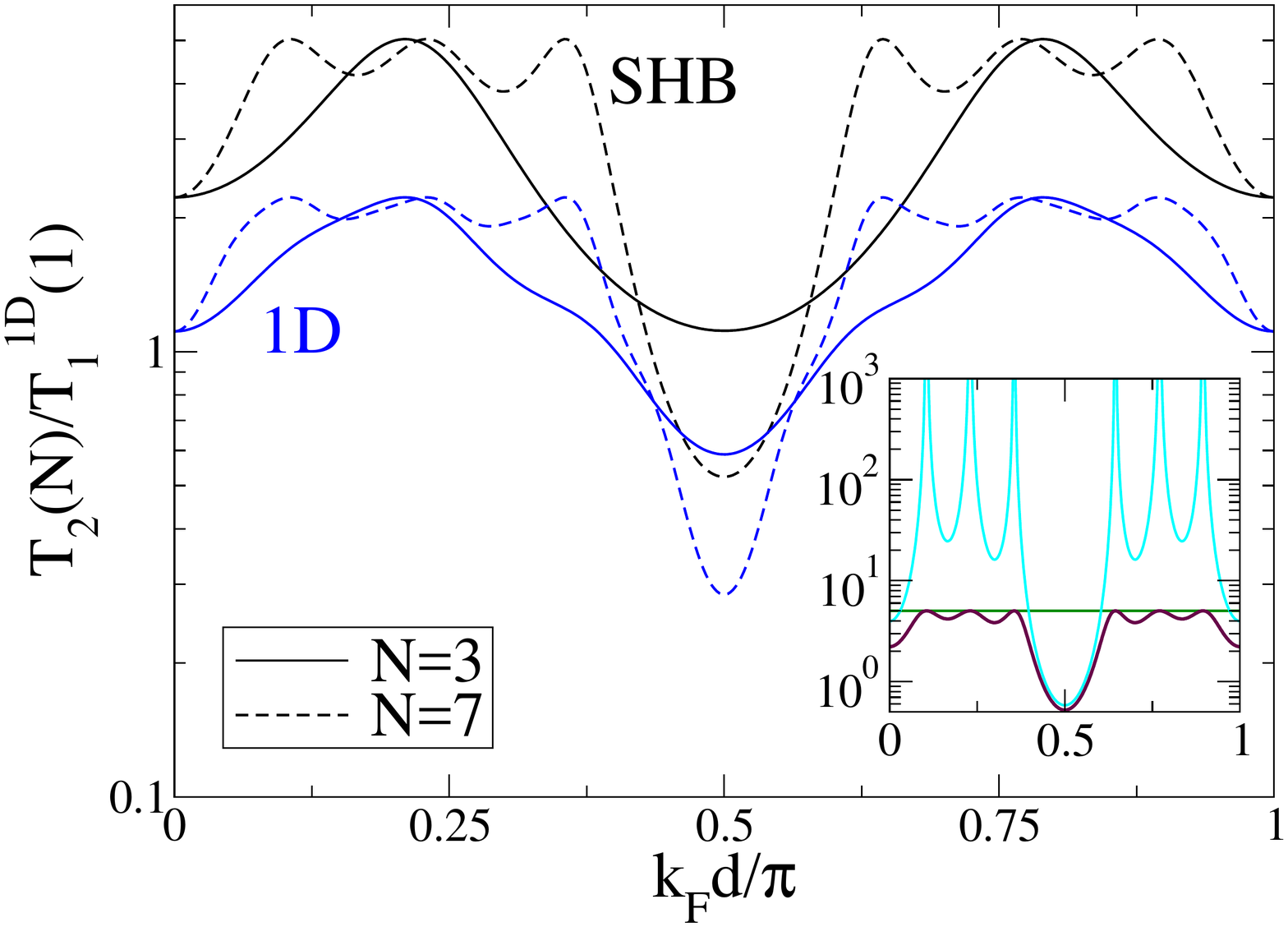}
\caption{ \label{fig4}
Spin decoherence times $T_2(N)$ for odd $N=3$ (solid lines) and $7$ (dashed lines) 
$S=1/2$ spin chains.  The  decoherence times obtained for the coupling with a SHB (one dimensional conductor) are depicted in black (blue). Same parameters than in Figure~\ref{fig3}. The inset shows the adiabatic  $1/\gamma^{\rm ad.}$ (dark green) and non-adibatic $1/\gamma^{\rm nonad.}$ (cyan) contributions to the total decoherence time $T_2(N)$ (maroon) for the $N=7$ spin chain. }
\end{figure}

Let us now consider $T_2$.
 The resulting decoherence times $T_2=1/(\gamma^{\rm ad.}+\gamma^{\rm nonad.})$ are depicted in Figure~\ref{fig4}. 
As observed, the SBH case lead to a decoherence time that is roughly a factor 2 larger than the coherence time of the chain coupled to a 1D electron gas. However, for most of the $k_Fd$ values, increasing the chain size (and thus, the relaxation times) does not lead to reduction of $T_2$.  The modulation of $T_2$ with $k_Fd$ is very mild, compared to the case of $T_1$, because the contribution of the spin-conserving forward scattering channel can not be eliminated.
This can be seen in the inset of Figure~\ref{fig4} where the adiabatic and non-adiabatic contributions to the decoherence time are depicted for the $N=7$ spin chain. As observed, the (momentum-independent) spin conserving contribution $1/\gamma^{\rm ad.}$, see equation (\ref{decoheb}) and table \ref{Table1}, remains always finite and it dominates the decoherence for most of the $k_Fd$ values.

Finally, we should notice that for both coupling to a 1D conductor or to a spin-momentum locked edge state, the relaxation and decoherence times can be substantially larger than for a normal three-dimensional metal with the same density of states at the Fermi energy and the same Kondo coupling~\cite{Delgado_Rossier_prb_2017}. 

\section{Discussion and conclusions\label{conclusions}}

In this work we have explored a new route to boost the spin lifetimes of magnetic atoms on surfaces.
To do so, we have taken advantage of the flexibility offered by spin chains by exploring the interference of the quasiparticle scattering at the different positions of the array, combined with the spin-momentum locking of the chiral edge states of a quantum spin Hall insulator.

In the case of atoms coupled to a normal conductor, spin lifetimes are mainly limited by the Kondo-induced spin-flip processes with quasiparticles at the Fermi surface of the conductor. Therefore, a clear way to increase the spin lifetimes is by reducing the number of scattering channels. This can be achieved for instance in a one-dimensional conductor, where the only allowed processes are forward and backward scattering. Yet, the spin-momentum locking at the borders of a quantum spin Hall conductor offers an additional thrilling feature: the forward spin-flip channel and the backward spin-conserving channel are fully quenched.

Experimental realizations of magnetic chains placed on the edges of a spin Hall bar are still to be demonstrated. Yet, there are a few substrates candidates such as the monolayer crystals of WTe$_2$~\cite{Fei_Palomaki_natphys_2017,Tang_Zhang_natphys_2017,Wu_Fatemi_science_2018} or  atomically thin islands of Bi$(111)$~\cite{Drozdov_Alexandradinata_natphys_2014,Schindler_Wang_natphys_2018}. The advances in STM manipulation and probe together with the STM-induced electron spin resonance makes us think that such an experimental setups could be realized in the near future.

Even if these system can be realized experimentally, there may be some effects preventing the observation of these long relaxation and decoherence times. For instance, we have not explored the possible issues associated with beyond-first neighbours interactions in the chain, and we overlooked the possible RKKY interaction~\cite{Zhou_Wiebe_natphys_2010}. In addition, we have neglected the role of electron-electron interactions in the SHB, that often has a strong influence in 1D systems.


One natural question that emerges when talking about magnetic atoms on surfaces is the role of magnetic anisotropy and the spin size. Most of the magnetic atoms used for assembly engineered spin structures do not behave as spin $S=1/2$ systems and they suffer some magneto-crystalline anisotropy~\cite{Hirjibehedin_Lutz_Science_2006,Loth_Bergmann_natphys_2010,Khajetoorians_Wiebe_science_2011,
Loth_Baumann_science_2012,Spinelli_Bryant_naturematerials_2014,Yang_Bae_prl_2017}. Although results were not shown, we have analyzed the effects of magnetic anisotropy on $S>1/2$ spin chains. We verified that for AFM spin chains that can be described as a two level system, large $T_1$ and $T_2$ times can still be reached by tuning $k_Fd$ on a QSH, although results are not as prominent as those of the $S=1/2$ Heisenberg chain.

To sum up,  spin lifetimes of a $S=1/2$ spin encoded on odd-numbered  antiferromagnetic $S=1/2$  Heisenberg chains  are limited due to Kondo interaction with a substrate.  We have shown that this unwanted effect can be dramatically reduced when the Kondo interaction occurs with conduction electrons in the spin-locked  edge states of a quantum Spin Hall insulator. This enhancement of the spin lifetimes can be  tuned  by varying the product $k_Fd$ of Fermi wavenumber $k_F$ and inter-spin distance $d$.We have shown that, for specific values of $k_F d$, the spin-flip backward scattering is also suppressed. 
 This behaviour is peculiar of the SHB edge states, and it stands on the locking of the spin and momentum degrees of freedom. Moreover, we demonstrated that although the resulting spin coherence time is limited by the unavoidable spin-conserving forward scattering, it can be still larger than the single spin decoherence time.

\ack
JFR acknowledges financial support by 
  FCT- The Portuguese Foundation for Science and Technology, under the project "PTDC/FIS-NAN/4662/2014" (016656), MINECO-Spain (Grant No. MAT2016- 78625-C2) and Generalitat Valenciana  Prometeo2017/139.
FD acknowledges financial support from Spanish Government
through grants MAT2015-66888-C3-2-R,  Basque Government, grant IT986-16 and Canary Islands program {\em Viera y Clavijo}  (Ref. 2017/0000231). FD acknowledges the hospitality of the Departamento de F\'isica Aplicada  of the Universidad de Alicante
\vspace{1cm}




\providecommand{\newblock}{}

\end{document}